\documentstyle[12pt]{article}
\begin{document}

\begin{titlepage}
\nopagebreak
\begin{flushright}

hep-th@xxx/9311035
\\
LPTENS--93/43
\\
November 1993
\end{flushright}

\vglue 2.5 true cm
\begin{center}
{\large\bf
ON AN EQUATION ASSOCIATED WITH THE CONTACT LIE ALGEBRAS}

\vglue 1 true cm
{\bf Mikhail V. SAVELIEV}\\
{\footnotesize Laboratoire de Physique Th\'eorique de
l'\'Ecole Normale Sup\'erieure\footnote{ On leave of absence from the Institute
for High Energy Physics, 142284, Protvino, Moscow region, Russia.},\\
24 rue Lhomond, 75231 Paris C\'EDEX 05, ~France.
}\\
\medskip
\end{center}

\vfill
\begin{abstract}
\baselineskip .4 true cm
\noindent
In the framework of a Lie algebraic approach we study a nonlinear equation
associated with the contact Lie algebra ${\bf K}K_m$. This algebra appears
to be relevant
for some solvable models of field theory and gravity in higher dimensions.
\end{abstract}
\vfill
\end{titlepage}
\baselineskip .5 true cm

In this note we study a nonlinear equation associated in the framework
of the group--algebraic approach
for nonlinear dynamical systems \cite{LS} with the contact Lie algebra ${\bf
K}K_m$ considered in the continuum Lie algebras formulation \cite{SV}.
It seems that in the same way as the area--preserving diffeomorphisms
on two--dimensional torus governs the symmetry of the self--dual Einstein
space with the rotational Killing vector, described by the so--called heavenly
equation \cite{BF}, the symmetry realised by the contact Lie algebras has a
relation to a number of field theory and gravity models in higher dimensions.

Let us first recall briefly the definition of contact Lie algebras following
the notations of ref. \cite{Fu}. The vector fields transforming the $1$-form
\begin{equation}
\sum_{j=1}^mx_jdx_{j+m}+dx_{2m+1}
\label{1}
\end{equation}
into a form  which differs from (\ref{1}) by the multiplication by a formal
power series, are called contact fields, and the corresponding Lie algebra is
denoted  ${\bf K}K_m$. Let $F_n={\bf K}[x_1,\cdots , x_n]$, $n=2m+1$, be the
space of the formal power series, and define on it the structure of a Lie
algebra,
\begin{equation}
{}[f, g]=-\hat{f}_xg+\frac{\partial f}{\partial x_n}(\hat{D}-2)g-
\frac{\partial g}{\partial x_n}(\hat{D}-2)f,
\label{2}
\end{equation}
where
\begin{equation}
\hat{f}_x\equiv \sum_{j=1}^m(\frac{\partial f}{\partial x_{j+m}}
\frac{\partial }{\partial x_j}- \frac{\partial f}{\partial x_j}
\frac{\partial }{\partial x_{j+m}}); \quad \hat{D}\equiv \sum _{A=1}^{2m}
x_A\frac{\partial}{\partial x_A}.
\label{3}
\end{equation}
The formula
\begin{equation}
f\rightarrow \hat{f}_x+(\hat{D}-2)f\frac{\partial }{\partial x_n}
\label{4}
\end{equation}
then sets the isomorphism $F_n\rightarrow {\bf K}K_m$. Being considered as a
continuum  ${\bf Z}$--graded Lie algebra ${\cal G}(E)=\oplus _{p\in {\bf Z}}
{\cal G}_p$, ${\bf K}K_m$ possesses the following defining relations for the
elements $X_{0, \pm 1}(f)$, $f\in E$, parametrising its local part ${\cal
G}_{-1}\oplus {\cal G}_0\oplus {\cal G}_{+1}$,
\begin{eqnarray}
& & [X_0(f), X_0(g)]=-X_0(\hat{f}_xg), \nonumber \\
& & [X_0(f), X_{\pm 1}(g)]=-X_{\pm 1}(\hat{f}_xg\pm g(\hat{D}-2)f),
\label{5} \\
& & [X_{+1}(f), X_{-1}(g)]=-X_0(\hat{f}_xg-(\hat{D}-4)(fg)). \nonumber
\end{eqnarray}
Now, let ${\cal M}$ be a two--dimensional manifold endowed with a complex
structure, and $A$ be a ${\bf K}K_m$--valued  $1$-form on ${\cal M}$. Any such
$1$-form generates a connection form  of some connection in the trivial fibre
bundle. We suppose that $A$ is flat, so that with a local coordinate $z$
in ${\cal M}$, $A=A_+dz+A_-d\bar{z}$, where $A_{\pm}$ are some mappings from
${\cal M}$ to ${\bf K}K_m$, satisfying the zero curvature condition
\begin{equation}
\partial _+A_--\partial _-A_++[A_+, A_-]=0.
\label{6}
\end{equation}
Here and in what follows $\partial _+\equiv \partial /\partial z, \partial _-
\equiv \partial /\partial \bar{z}$. Choosing some basis in ${\bf K}K_m$ and
considering the components of the decomposition of $A_{\pm}$ over this basis as
fields, we can treat (\ref{6}) as a nonlinear system of partial differential
equations for the fields. In accordance with \cite{LS}, to provide
nontriviality
of such a system, we impose the so--called grading condition on the connection,
so that the $(1,0)$-component $A_+$ of $A$ takes values in
$\oplus _{p\geq 0}{\cal G}_{+p}$, and the $(0,1)$-component $A_-$ takes values
in $\oplus _{p\geq 0}{\cal G}_{-p}$. Moreover, we confine ourselves to the
case when $A_{\pm}$ take values
only in the local part of the algebra, i.e., in ${\cal G}_0\oplus {\cal
G}_{\pm 1}$, respectively. Due to the arbitrariness related to the gauge
transformations generated by ${\cal G}_0$, we can finally take
\begin{equation}
A_+=X_{+1}(f),\qquad A_-=X_0(u)+X_{-1}(g);\qquad u, f, g\in E.
\label{7}
\end{equation}
Then it is easy to convince oneself that a ${\cal G}_0$--gauge invariant
equation
arising from (\ref{6}) with (\ref{7}), can be represented in the form
\begin{equation}
\partial _+[(1-\hat{L})^{-1}\partial _-\rho ]=\frac{1}{2}(4-\hat{D})e^{\rho },
\label{8}
\end{equation}
where $2\hat{L}\equiv \hat{D}-\hat{\rho }_x$. Note that here $\rho (z,\bar{z};
x_1,\cdots , x_{2m})\equiv \mbox{ log }4f$, $u=-1/2(1-\hat{L})^{-1}\partial _-
\rho$, while the function $g$ can be taken to be unity, up to inessential
transformations. Using the formula
\[
\partial \frac{1}{1-\hat{L}}=\frac{1}{1-\hat{L}}\partial \hat{L}\frac{1}{1-
\hat{L}},\]
one can rewrite equation (\ref{8}) as
\begin{equation}
\partial _+\partial _-\rho+\partial _+\hat{L}(1-\hat{L})^{-1}\partial _-\rho
=\frac{1}{2}(1-\hat{L})(4-\hat{D})e^{\rho }.
\label{9}
\end{equation}
This equation looks quite complicated, however, the fact that the
algebra ${\bf K}K_m$ is of the finite growth, see e.g. \cite{Kac}, allows us to
believe that it can be integrated in this or that sense.

Let us consider special simplifications of the equation in question, based
on some additional symmetry conditions imposed on it. It is clear that if the
function $\rho $ depends only on $z, \bar{z}$, and the ratios $x_j/x_{j+m}$,
then equation (\ref{9}) is automatically reduced to the Liouville equation,
$\partial _+\partial _-\rho _L=2e^{\rho _L}$. On the other hand, if $\rho =
\rho (z,\bar{z}; r\equiv -\mbox{ log }\sum _Ax_A^2)$, then we arrive at
the form
\begin{equation}
\partial _+\partial _-\rho = (2+3\partial /\partial r +\partial
^2/\partial r^2)e^{\rho }.
\label{10}
\end{equation}
This equation evidently possesses a self--similar solution
\begin{equation}
e^{\rho }=e^{\rho _L(z,\bar{z})}(1+ \alpha_1e^{-r}+\alpha_2e^{-2r})
\label{11}
\end{equation}
with arbitrary constants $\alpha_{1,2}$. It plays the same role
as the Eguchi--Hanson gravitational instanton for the self--dual Einstein
space with the rotational Killing vector, described by the completely
integrable equation $\partial _+\partial _-\rho _H=\partial ^2/\partial r^2
e^{\rho _H}$, and related to the continuum Lie algebra ${\cal
G}(E;K,\mbox{id})$
with the Cartan operator $K=\frac{\partial ^2}{\partial r^2}$, which in turn
is isomorphic to $S_0\mbox{ Diff }T^2$.

Equation (\ref{10}) is written in a form of the continuous Toda
system with the Cartan operator $K=\frac{\partial
^2}{\partial r^2}+3\frac{\partial }{\partial r}+2$, and, correspondingly,
is associated with the continuum Lie algebra ${\cal G}(E;K,\mbox{id})$, to
which the algebra ${\bf K}K_m$ is reduced in this case. It is interesting
to note that ${\cal G}(E; \frac{\partial ^2}{\partial r^2}+3\frac{\partial
}{\partial r}+2, \mbox{id})$ realises a special case $\{c_0=c_1\not= 0, c_2
=0\}$ of the Lie algebra $W^{(c_0,c_1,c_2)}$ introduced in
\cite{KSSV}, see also \cite{Hop}, as a
modification of $S_0\mbox{ Diff }T^2$, with the elements satisfying,
in a component form, the commutation relations
\[
{}[Y_{\bf m}, Y_{\bf n}]=(c_0{\bf m}\times {\bf n}+{\bf c}({\bf m}-{\bf n}))
Y_{{\bf m}+{\bf n}} + \mbox{ central term }.\]
Here ${\bf m}=(m_1,m_2)$, ${\bf n}=(n_1,n_2)$ are two--dimensional integer
vectors, ${\bf m}\times {\bf n}\equiv m_1n_2-m_2n_1$; and ${\bf c}=(c_1,c_2)$
is a constant vector.

The role of the (proper) B\"acklund map for equation (\ref{10}) for the case
when the function $\rho $ depends not on both variables $z$ and $\bar{z}$, but
on their linear combination, say $t\equiv \frac{z-\bar{z}}{\sqrt{2}}$, is
played by the equation
\begin{equation}
\frac{\partial \rho}{\partial t} = (1+\frac{\partial}{\partial r})e^{\rho /2},
\label{12}
\end{equation}
which coincides with the classical equation for the Riemann reversal
wave written in terms the function $\Phi (t, \tau \equiv 2e^r)=\frac{\tau}{2}
e^{\rho (t, \tau )/2}$. Just the overlap phenomena caused by the equation for
$\Phi$ can lead in turn, in the same way as for the continuous long wave
approximation of the Toda system (the heavenly equation), see e.g. \cite{Lax},
to the generation of shock waves.

There are some other symmetry conditions, e.g. also of a scaling type, which
considerably simplify equation (\ref{9}).

Finally, let us note that the system under consideration, described by
equations (\ref{8}) or (\ref{9}), can be naturally presented in a symmetrical
form with respect to the coordinates $z_+$ and $z_-$, namely
\begin{eqnarray}
\partial _+[(1-\hat{L}_-)^{-1}\partial _-\rho ] +
\partial _-[(1-\hat{L}_+)^{-1}\partial _+\rho ] & + &
\frac{1}{4}\{(1-\hat{L}_-)^{-1}\partial _-\rho ,
(1-\hat{L}_+)^{-1}\partial _+\rho \}_P \nonumber \\
& = & (4 - \hat{D})e^{\rho },
\label{13}
\end{eqnarray}
where
\[
2\hat{L}_{\pm}\equiv \hat{D}\pm \frac{1}{2}\hat{\rho }_x; \qquad
\{f, g\}_P \equiv \hat{f}_x g.\]
Moreover, in the case when $\rho =
\rho (t; x_1,\cdots , x_{2m})$, this equation can be written
in the Brockett double commutator form \cite{Br} which is relevant for a
description of some sorter (analog) systems.

\bigskip

Finishing up the paper, I would like to thank J.-L. Gervais and A. M. Vershik
for the illuminating discussions, and
the Laboratoire de Physique Th\'eorique de l'\'Ecole Normale Sup\'erieure
de Paris for kind hospitality and excellent conditions for my scientific
work here.


\begin{thebibliography}{**}
\bibitem{LS}
A. N. Leznov and M. V. Saveliev, {\it Group--Theoretical Methods for
Integration
of Nonlinear Dynamical Systems}, Progress in Physics Series, v. 15
(Birkha\"user--Verlag, Basel, 1992).
\bibitem{SV}
M. V. Saveliev and A. M. Vershik, Comm. Math. Phys., {\bf 126} (1989) 367.
\bibitem{BF}
C. Boyer and D. Finley, J. Math. Phys., {\bf 23} (1982) 1126.
\bibitem{Fu}
D. B. Fuks, {\it Cohomology of infinite--dimensional Lie algebras},
(Contemporary Soviet Mathematics, Consultants Bureau, New York, 1986).
\bibitem{Kac}
V. G. Kac, {\it Infinite Dimensional Lie Algebras}, 2nd ed.,
(Cambridge University Press, Cambridge, 1985).
\bibitem{KSSV}
R. M. Kashaev, M. V. Saveliev, S. A. Savelieva and A. M. Vershik, in: {\it
Ideas and Methods in Mathematical Analysis, Stochastics, and Applications},
eds. S. Albeverio, J. E. Fenstad, H. Holden and T. Lindstrom, Vol. 1
(Cambridge University Press, Cambridge, 1992).
\bibitem{Hop}
J. Hoppe, Rev. Math. Phys., {\bf 2}, \#2 (1990) 193.
\bibitem{Lax}
P. Lax, Comm. Pure Appl. Math., {\bf 44} (1991) 1047.
\bibitem{Br}
R. W. Brockett  and A. Bloch, in: {\it Proc. CRM Workshop on Hamiltonian
Systems, etc.}, eds. J. Harnad and J. E. Marsden (Montreal, 1990).
\end{thebibliography}
\end{document}